# Distributed delay-line interferometer based on a Bragg grating in transmission mode

**MIGUEL A. PRECIADO,[1,2] ATALLA EL-TAHER[1], XUEWEN SHU[1,3,*] AND KATE SUGDEN[1]**

[1] *Aston Institute of Photonic Technologies, Aston University, Birmingham B4 7ET, UK*

[2] *School of Physics and Astronomy, University of Glasgow, Glasgow G12 8QQ, UK*

[3] *Wuhan National Laboratory for Optoelectronics, Huazhong University of Science and Technology, Wuhan, China.*

[*] *corresponding author: xshu@hust.edu.cn*

**A novel approach for a delay line interferometer (DLI) based purely on forward Bragg scattering is proposed. We have numerically and experimentally demonstrated that a Bragg grating can deliver the functionality of a DLI in its transmission mode along a single common interfering optical path, instead of the conventional DLI implementation with two interfering optical paths. As a proof of concept, a fiber Bragg grating has been designed and fabricated, showing the desired functionality in the transmission mode of the Bragg grating. The proposed "Bragg-DLI" approach is applicable to any kind of Bragg grating technology, such as volume Bragg gratings, dielectric mirrors, silicon photonics, and other optical waveguide based Bragg structures.**

## INTRODUCTION

Delay line interferometers (DLI), such as Mach-Zehnder or Michelson-Morley interferometers, are basic optical devices where the output optical signal is generated by interfering two replicas of the input signal, commonly generated by splitting the input signal in two different optical paths and re-combining them to generate a resulting interference signal. DLIs are used in a wide range of applications, such as the characterization of optical sources: a number of optical signal processing applications: QPSK, DPSK and PSK demodulator and data conversion [1,2]; photonic quantization [3]; chirp and linewidth characteristics of semiconductor lasers [4]; and sensing of various physical and chemical parameters such as temperature, strain, refractive index, displacement curvature, inclination, or vibration [5-12].

Here we report a novel patented [13] implementation of a DLIs based on a Bragg grating (BG) operated in transmission mode, with a fundamentally different physical working principle from conventional DLIs, illustrated in Fig. 1. Instead of having two physically separated optical paths for the interfering signals as showed in Fig. 1(a), in the proposed Bragg-DLI the interfering signals are simultaneously generated and interfered by forward Bragg scattering, in a single common path along the BG, represented in Fig. 1(b). As a proof of concept, we present the design and fabrication of a BG-DLI using an in-fiber implementation, concretely a phase-modulated fiber BG (FBG) in transmission mode, showing the DLI functionality of the device in numerical simulations and experimental measurements. Finally, we conclude the Letter with a summary and discussion of the results.

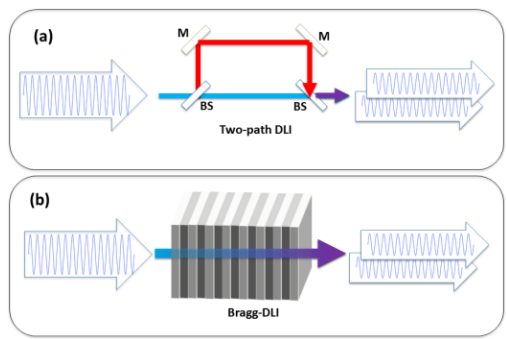

Fig. 1. Comparison of a conventional two-path DLI (free-optics implementation) (a), and a single-path Bragg-DLI (b). BS: Beam Splitter; M: Mirror.

## DESIGN AND NUMERICAL RESULTS

An ideal DLI functionally can be expressed in the temporal domain as

$$f_{out}(t) \propto f_{in}(t) + f_{in}(t-T)\exp(j\phi) \qquad (1)$$

where $f_{in}(t)$ and $f_{out}(t)$ are the complex envelopes of the input and output signals respectively, $\phi$ and $T$ are the relative phase and delay between the interfering components. Equivalently in the frequency domain, we can obtain the spectral transfer function, $H(\omega) = F_{out}(\omega) / F_{in}(\omega)$, as

$$H(\omega) \propto 2\cos(\omega T/2 - \phi/2)\exp(-j\omega T/2 - j\phi/2) \qquad (2)$$

, where $F_{out}(\omega)$ and $F_{in}(\omega)$ are the Fourier transforms of $f_{out}(t)$ and $f_{in}(t)$, respectively, $\omega$ is the base-band angular pulsation i.e. $\omega = \omega_{opt} - \omega_0$, $\omega_{opt}$ is the optical angular pulsation, $\omega_0$ is the central angular frequency of the signals, and $j=(-1)^{1/2}$ is the imaginary unit.

BGs operating in transmission are minimum phase systems [14], therefore it is not possible to achieve a completely arbitrary response using these photonic structures. Fortunately, in our particular case, the DLI transfer function satisfies the minimum phase condition [15], HT{log|$H(\omega)$|}=∠{$H(\omega)$}, where HT denotes the Hilbert transform operator, and ∠ denotes the phase operator, which implies that when a BG is designed for delivering the DLI spectral response amplitude in its transmission mode described in Eq. (2), the DLI spectral response phase, and therefore the temporal response described in Eq. (1), is also automatically obtained.

In principle, there is no theoretical restriction for the kind of BG technology used in the physical implementation of the BG-DLI. Here, we have used a fiber Bragg grating (FBG) implementation operating in transmission mode [14,16-25] in order to demonstrate the proposed approach. Phase-modulated FBGs, initially proposed for virtual Gires-Tournois interferometers in reflection mode [26], have also been proposed and numerically demonstrated as an alternative feasible implementation for some specific spectral responses for pulse shaping in transmission mode [25]. The grating strength is more challenging to accurately control in the fabrication process than the grating period due to the fiber photosensitivity variability, and the non-linear relation between the writing illumination power and the corresponding fiber core refractive index modification. However, in a phase-modulated BG the grating strength remains basically uniform along most of the grating length, while the grating functionality is defined by modulating the grating period, which is much easier to accurately control in the fabrication process in practice.

In this design process we have taken into account the capabilities of our fabrication system. We have defined a grating length of 9 cm with a bandwidth of approximately 4 nm centered at $\lambda_0$=1550 nm. Following a numerical optimization process using a similar procedure to that described in [25], we have obtained a grating period function corresponding to the desired spectral function $H(\omega)$, where the DLI parameters have been set to $T$=20 ps, and $\phi$=0. The resulting grating strength and grating period is shown in Fig. 2 (a), and the corresponding data set is accessible in **Data File 1** to facilitate the reproducibility of these results to other researchers. As it can be observed in Fig. 2 (b), the numerically simulated spectral response in transmission of the FBG is in good agreement with the ideal response of the DLI.

In order to validate the functionality of the designed BG-DLI, we define two examples. The first one with a single pulse as an input signal to show the basic working principle, and the second one with a test sequence of pulses as an input signal that includes all possible cases of the DLI functionality. Let us define an input signal compose by a single optical pulse $f_{in}(t) = p(t)$ for the first example, where $p(t)$ is the complex envelope function of optical Gaussian pulses of 5 ps full width half maximum (FWHM), with central wavelength $\lambda_0=2\pi/\omega_0$=1550 nm. From Eq. (1) we can deduce that the output sequence of the ideal DLI is $f_{out}(t) = \sum_{k=1}^{2} d_k p(t - kT)$, with $d_k \propto \{1,1\}$ (i.e. $d_1 \propto 1$, $d_2 \propto 1$), with an intensity $I_{out}(t) = |f_{out}(t)|^2 \simeq \sum_{k=1}^{2} |d_k|^2 |p(t-kT)|^2$, where we have assumed $p(t-mT)p^*(t-nT)\simeq 0$ for m≠n (negligible pulses overlapping), and the total transmission delay is neglected. As it can be observed in Fig. 3, the output intensity and phase obtained from numerical simulations of the designed BG from the previously defined $f_{in}(t)$ is in good agreement with $I_{out}(t)$.

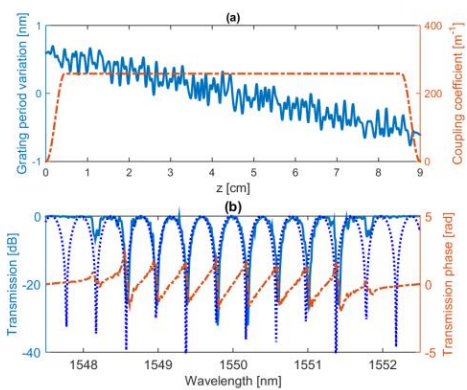

Fig. 2. (a) Grating apodization (red-dash-dotted) and period (blue-solid) obtained from numerical optimization to approach the desired spectral response. (b) Numerically simulated spectral response amplitude of the designed FBG in transmission mode, obtained in amplitude (blue-solid), and phase (red-dash-dotted), and ideal spectral response amplitude (blue-dotted) used as objective in the numerical optimization.

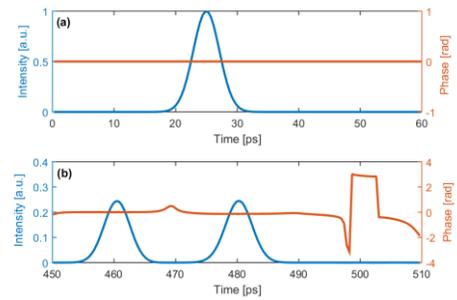

Fig. 3. Optical intensity and phase of input signal (a), and output signal (b) resulting from numerical simulations of the designed grating for the first example (single pulse input), where the basic functionality of the BG-DLI is shown.

In the second example we define a test sequence composed of multiple optical pulses to numerically verify the BG-DLI functionality. The pulses sequence is defined by $f_{in}(t) = \sum_{k=1}^{5} c_k p(t - kT)$, where $c_k$={1,0,1,1,-1}. Applying Eq. (1) we can deduce that the output sequence of the ideal DLI is given by: $f_{out}(t) = \sum_{k=1}^{5} d_k p(t - kT)$, where $d_k \propto c_k + c_{k+1}$={1,1,1,2,0,-1}, and with an intensity of $I_{out}(t) = |f_{out}(t)|^2 \simeq \sum_{k=1}^{5} |d_k|^2 |p(t-kT)|^2$, where again we have assumed $p(t-mT)p^*(t-nT)\simeq 0$ for m≠n (negligible pulses overlapping), and the total transmission delay is not taken into account. In Fig. 4, we can observe a good agreement the numerically obtained output intensity, and the expected $I_{out}(t)$ previously defined.

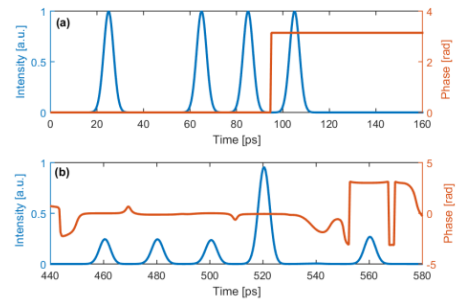

Fig. 4. Optical intensity and phase of input signal (a), and output signal (b) resulting from numerical simulations of the designed grating for the second example (multiple pulses input), where all possible cases of pulses interferences are showed.

In order to illustrate the physical generation of the interference of the BG-DLI in a common interfering path, and to emphasize the difference with conventional two interfering paths DLIs approaches, we have also calculated the optical intensity distribution inside the designed BG as it propagates along the grating by using the numerical method proposed in [27] by Muriel et al. As it can be observed in Fig. 5(a) for the first example, the incident pulse is temporally separated into two components as the signal propagates along the common path, without requiring a spatial splitting in different paths at any point. For the case of the pulse sequence represented in Fig. 5(b), corresponding to the second example, the previous process is repeated for each pulse of the input sequence, and the overlapping of the resulting separated pulses leads to pulses interferences (constructive or destructive) in the output signal in a distributed way along the grating.

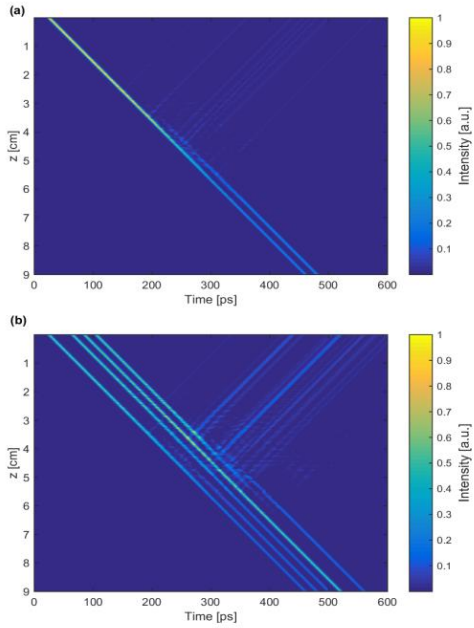

Fig. 5. Optical intensity distribution inside the FBG for the first example (a), and the second example (b). See **Visualization 1** for a detailed observation of the spatial distribution of the optical intensity inside the fiber as a function of time.

## EXPERIMENTAL RESULTS

A phase-modulated FBG has been fabricated with the previously calculated grating strength and period, using a UV laser direct-writing system developed at Aston University, where the grating is created pitch-by-pitch, and the coupling coefficient profile and the varied period are realized by controlling the ON/OFF of an acoustic optical modulator and moving the phase mask/fiber. A hydrogen-loaded photosensitive fiber, later stabilized by annealing at 80°C for 60 h after the fabrication of the grating, was used in the process. The fabricated FBG has been characterized using a broad spectrum light source as excitation, and measuring the output spectrum of the FBG in transmission, which is shown in Fig. 6(a) compared to the ideal DLI interferometer. The corresponding spectral phase is also showed in Fig. 6(b), which has been numerically recovered from the previous spectral intensity measurement by using the Hilbert transform relations of the transmission spectral response amplitude/phase [14], a method reported as more robust than measurement by spectral interferometry in [28]. As it can be observed, the errors in the fabrication process affects as a distortion of the FBG spectral response amplitude and phase observed in Fig. 6 (a) and (b), as it is described in [29]. However, it is worth noting that the transmission mode phase response is less sensitive to grating-fabrication errors than in reflection mode [14,30]. In any case, a reasonably good agreement between ideal and fabricated DLI response can be observed in the bandwidth of interest for both amplitude and phase spectral response.

The response to optical pulses in transmission mode of the fabricated FBG has been characterized by using a pulsed laser based on an Erbium gain based single-walled carbon nanotube (CNT) passively-mode-locked fiber laser as excitation, where the spectrum of these pulsed laser input and the corresponding FBG-DLI output is also showed in Fig. 6(a). The input and output signals has been also characterized in temporal domain by using an intensity auto-correlator, where the autocorrelation of input and output is showed in Fig. 6(b). As it can be observed, the autocorrelation intensity indicates an output signal composed by a double pulse with a relative delay of approximately 20 ps in the output signal, matching the designed DLI functionality.

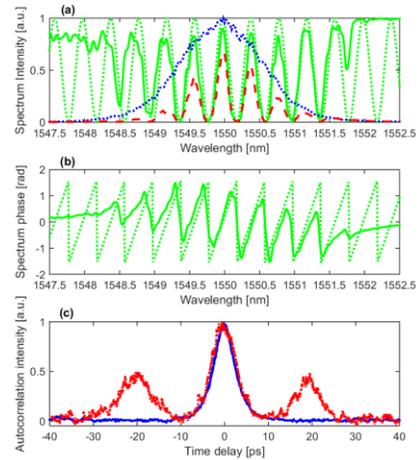

Fig. 6. (a) Spectral response intensity of the ideal DLI (green-dashed), compared to the measured spectral response intensity (green-solid) of the fabricated FBG in transmission when excited with a broad source; spectrum of the pulsed laser source used as an input signal (blue-solid), and the corresponding output of the FBG in transmission mode (red-dashed);(b) Spectral response phase of the ideal DLI (green-dashed) compared to the Hilbert transform recovered spectral response phase of the fabricated FBG (green-solid) ;(c) auto-correlation trace for input (blue-solid) and output (red-dotted) signals, revealing a double pulse output separated 20 ps (as designed) from a single pulse input.

Regarding the environmental sensitivity/tunability properties of the fabricated DLI-FBG, we have verified that similar to other FBGs [31-33] the central Bragg wavelength can be shifted by the application of either temperature or strain variation while the spectral response remains approximately undistorted. In Fig. 7, dependence to temperature and strain is showed, where we obtained a spectral shift of $9.81 \times 10^{-3}$ nm/°C for temperature, and $1.14 \times 10^{-3}$ nm/µε for strain. Taking into account that a shift in wavelength of 0.8014 nm (free spectral range) corresponds to a variation of $2\pi$ in $\phi$, we can easily deduce the dependence of $\phi$ on temperature to be 76.9 mrad/°C, and on strain to be 8.9 mrad/µε. These values are significantly smaller than the typically very high sensitive two-path DLIs, where a relative variation of optical paths equivalent to a wavelength produces a variation of $2\pi$ in $\phi$

A proper package is needed when a reduced environmental sensitivity of an optical device is required. An athermal packaged FBG reported in [34] can reduce the thermal sensitivity to a value of $0.41 \times 10^{-3}$ nm/°C, which corresponds to a shift in $\phi$ of 3.2 mrad/°C. Most recent commercial packaging report an even lower thermal sensitivity of < 0.01 nm (*OEquest*, part number 91000223-063), in a temperature range from -5 °C to +70 °C, which correspond to maximum DLI relative phase variation $\phi$ <78.4 mrad in this temperature range of 75 °C . In [35] a packaged all fiber MZI*s* ,with same delay amount as our fabricated DLI of 20 ps, is reported with a thermal shift in $\phi$ of 91.9 mrad/°C, which is significantly higher than that of a packaged FBG (even also higher when compared to the unpackaged FBG).

These properties suggest a possible use of proposed BG-DLI when low sensitivity is desired, such as signal processing applications where environmental perturbations are undesirable, or high range sensing applications, possibly combined with a more sensitive sensor of smaller range.

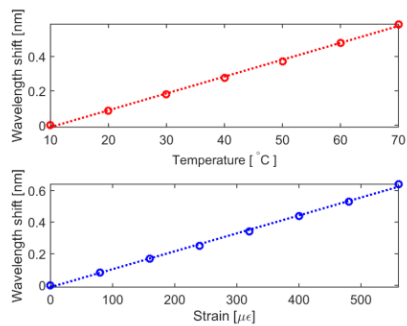

Fig. 7. Spectral shift of the transfer function of the fabricated fiber Bragg-DLI due to temperature variation (a) and strain (b).

## CONCLUSION AND DISCUSSION

We have proposed and demonstrated a novel implementation of a DLI based on a Bragg resonant structure in transmission mode, with a fundamental working principle radically different to that of conventional two-interfering paths DLIs. As a proof of concept, an optical fiber implementation of a Bragg-DLI using a phase-modulated FBG was designed and fabricated. In the numerical simulations we have shown the desired DLI functionality of the designed FBG, and for illustrative purpose the optical intensity distribution inside the Bragg grating has also been calculated, showing how the forward Bragg scattering generates a single common path interference in a distributed way along the Bragg grating. An FBG has been fabricated according to the design parameters, and the temporal and spectral characterization shows a good agreement with the desired DLI functionality. It is worth noting that in this Letter we also report the first experimental demonstration of a phase-modulated FBG for the synthesis of a specific spectral response in transmission mode, which were proposed but only numerically demonstrated in [25] for some pulse shaping applications.

Sensitivity to environmental changes is typically very high in conventional two path based DLIs, where any relative variation of the effective optical paths of the interfering components in the order of a fraction of a wavelength produces a significant variation of the response, making them very suitable for sensing applications of small perturbations. Additionally, for in-fiber implementation of two-paths DLIs, any relative difference in the polarization state in the DLI paths will affect to the performance.

In our case, the proposed Bragg-DLI have a common interfering path, showing a higher robustness to environmental changes, which may be more indicated for applications where a well-controlled DLI operation is desired, or for sensing a higher range of variations of environmental variables. The central wavelength can be accurately tuned by strain or temperature similarly to conventional FBG, while the whole response remains basically undistorted. It is worth noting that only $\phi$ can be tuned in practice in the FBG-DLI, while the relative delay $T$ remains approximately constant. It is worth noting that this limitation also applies to most conventional waveguide MZIs, where the induced variations in the optical path by application of temperature in one of the MZI arms only affects significantly to the relative phase, but very marginally to the relative delay.

In principle, this approach can be implemented in any kind of Bragg grating technology, such as volume Bragg gratings, dielectric mirrors, silicon photonics, or other optical waveguide technologies based Bragg structures, introducing an alternative for the physical implementation DLIs based in a common optical interfering path.

**Funding.** Marie Curie IEF, FP7-PEOPLE-2010-IEF-275703-IFOCS.

## SUPPLEMENTARY MATERIALS

**Dataset 1.** https://doi.org/10.5281/zenodo.167590

**Visualization 1.** https://doi.org/10.5281/zenodo.167588

## REFERENCES

1. R. Kou, H. Nishi, T. Tsuchizawa, H. Fukuda, H. Shinojima, and K. Yamada, Opt. Express 20, 11037-11045 (2012).
2. Z. Zhang, Y. Yu, and X. Zhang, Opt. Express 19, 12427 (2011).
3. H. Chi, Z. Li, X. Zhang, S. Zheng, X. Jin, and J. Yao, Opt. Lett. 36, 1629 (2011).
4. N. Satyan, A. Vasilyev, G. Rakuljic, V. Leyva, and A. Yariv, Opt. Express 17, 15991 (2009).
5. L. Xu, Y. Li, and B. Li, Appl. Phys. Lett. 101, 153510 (2012).
6. Y. Xu, Ping Lu, Zengguang Qin, Jeremie Harris, Farhana Baset, Ping Lu, Vedula Ravi Bhardwaj, and Xiaoyi Bao, Opt. Express 21, 3031 (2013)
7. H. F. Martins, J. Bierlichc, K.Wondraczekc, S. Ungerc, J. Kobelkec, K. Schusterc, M. B. Marquesa,b, M. Gonzalez-Herraezd, O. Frazãoa, OFS2014 23rd International Conference on Optical Fiber Sensors. International Society for Optics and Photonics, 2014.
8. P. Lu and Q. Chen, Opt. Lett. 36, 268 (2011).
9. Z. Tian, S. S.-H. Yam, J. Barnes, W. Bock, P. Greig, J. M. Fraser, H.-P. Loock, and R. D. Oleschuk, IEEE Photon. Technol. Lett. 20, 626 (2008).
10. J. Chen, J. Zhou, and X. Yuan, IEEE Photon. Technol. Lett. 26, 837 (2014).
11. T. Wei, X. Lan, and H. Xiao, IEEE Photon. Technol. Lett. 21, 669 (2009).
12. C. R. Liao, Y. Wang, D. N. Wang, and M. Yang, Fourth European Workshop on Optical Fibre Sensors. International Society for Optics and Photonics (2010).
13. X Shu, MA Preciado, US Patent 20160109657.
14. J. Skaar, J. Opt. Soc. Am. A 18, 557 (2001).
15. F.W. King, *Hilbert Transforms* (Cambridge University Press, 2009).
16. N. M. Litchinitser, B. J. Eggleton, and D. B. Patterson, J. Lightwave Technol. 15, 1303 (1997).
17. M. A. Preciado and M. A. Muriel, Opt. Lett. 33, 2458 (2008)
18. L. M. Rivas, S. Boudreau, Y. Park, R. Slavík, S. LaRochelle, A. Carballar, and J. Azaña, Opt. Lett. 34, 1792 (2009).
19. N. Q. Ngo, Opt. Lett. 32, 3020 (2007)
20. M. H. Asghari and J. Azaña, Opt. Express 16, 11459(2008).
21. M. A. Preciado and M. A. Muriel, Opt. Lett. 34, 752 (2009).
22. M. Fernández-Ruiz, M. Li, M. Dastmalchi, A. Carballar, S. LaRochelle, and J. Azaña, Opt. Lett. 38, 1247-1249 (2013).
23. M. Burla, M. Li, L. Cortés, X. Wang, M. Fernández-Ruiz, L. Chrostowski, and J. Azaña, Opt. Lett. 39, 6241-6244 (2014).
24. M. Fernández-Ruiz, L. Wang, A. Carballar, M. Burla, J. Azaña, and S. LaRochelle, Opt. Lett. 40, 41-44 (2015).
25. M. Preciado, X. Shu, and K. Sugden, Opt. Lett. 38, 70-72 (2013).
26. X. Shu, K. Sugden, and K. Byron, Opt. Lett. 28, 881-883 (2003)
27. M. A. Muriel and A. Carballar, IEEE Photon.Technol. Lett., vol. 9, no. 7, p. 955, Jul. 1997.
28. M. Dicaire, J. Upham, I. De Leon, S. A. Schulz, and R. W. Boyd, J. Opt. Soc. Am. B 31, 1006-1010 (2014)
29. R. Feced, M. N. Zervas, J. Lightwave Technol. 18, 90-101 (2000).
30. K. Hinton, J. Lightwave Technol. 16, 2336- (1998)
31. K.-M. Feng, J.-X. Cai, V. Grubsky, D. S. Starodubov, M. I. Hayee, S. Lee, X. Jiang, A. E. Willner, and J. Feinberg, IEEE Photon. Technol. Lett., vol. 11, pp. 373–375, 1999.
32. C. S. Goh, S. Y. Set, and K. Kikuchi, IEEE Photon. Technol. Lett., vol. 14, no. 9, pp. 1306–1308, Sep. 2002
33. J. Lauzon, S. Thibault, M. J, and F. Ouellette, Opt. Lett., vol. 19, pp. 2027–2029, 1994.
34. Y. Huang, J. Li, G. Kai, S. Yuan, and X. Dong, Microwave and Optical Technology Letters, vol. 39, no. 1, pp. 70-72 (2003).
35. G. Ducournau, O. Latry, and M. Ketata, Syst., Cybernet. Inf., vol. 4, no. 4, pp. 78–89, 2006.